\documentclass[manuscript]{aastex}

\begin{document}

   \title{Apparent Trend of the Iron Abundance in NGC 3201: the Same Outcome with Different Data}

\author{Valery V. Kravtsov}
\affil{Departamento de F\'isica, Facultad de Ciencias Naturales, Universidad de Atacama, Copayapu 485, Copiap\'o, Chile\\
              Sternberg Astronomical Institute, Lomonosov Moscow State University, University Avenue 13,
              119899 Moscow, Russia\\}
\email{valery.kravtsov@uda.cl}

\begin{abstract}
We further study the unusual trend we found at statistically
significant levels in some globular clusters, including NGC 3201: a
decreasing iron abundance in red giants towards the cluster centers.
We first show that recently published new estimates of iron
abundance in the cluster reproduce this trend, in spite of the
authors' statement about no metallicity spread due to a low scatter
achieved in the [FeII/H] ratio. The mean of [FeII/H] within
$R´\sim2\arcmin$ from the cluster center is lower, by
$\Delta$[FeII/H] = 0.05$\pm$0.02 dex, than in the outer region, in
agreement with our original estimate for a much larger sample size
within $R \approx9\arcmin$. We found that an older dataset traces
the trend to a much larger radial distance, comparable with the
cluster tidal radius, at $\Delta$[Fe/H]$\sim$0.2 dex due to higher
metallicity of distant stars. We conclude the trend is reproduced by
independent datasets and find that it is accompanied with both a
notable same-sign trend of oxygen abundance which can vary by up to
$\Delta$[O/Fe]$\sim$0.3 dex within $R \approx9\arcmin$, and
opposite-sign trend of sodium abundance.

\end{abstract}


\maketitle

\section{Introduction}
\label{introduc}

Analyzing archival data of elemental abundances in red
giants (RGs) of a sample of so-called monometallic globular clusters
(GCs), we found \citep{kravtsov13} radial trend of iron abundance,
statistically significant at high and very high confidence levels,
in a number of the sample GCs. This trend is fairly unusual given
(1) the fact itself of its existence in the GCs and (2) that the
[Fe/H] ratio is lower, by about $\Delta$[Fe/H]$\sim$0.05 dex, in the
central than in the outer parts of the GCs. The key points to
understand are: (1) whether the trend
is reproduced in independently obtained  data sets on iron
abundance in the same GCs; and, if it is the case, (2) whether it is
caused by the real radial variation of iron abundance or by any kind
of radial systematic effect mimicking it. The first point implies a
need to study various data originating from observations gathered
from different facilities and/or reduced independently.

In our original work \citep{kravtsov13} we used fairly uniform data
on spectroscopy of large samples of RGs in a diversity of globular
clusters GCs (not less than 100 stars in each of the GC sample: NGC
104, NGC 288, NGC 1851, NGC 3201, NGC 4590, NGC 5904, NGC 6121, NGC
6254, MGC 6752, and NGC 6809), obtained by
\citet{carrettal07,carrettal09,carrettal11} using the
same facility, namely the VLT FLAMES-GIRAFFE spectrograph ($R=22,500$). For a
sample of RGs in each GC, we compared the radial distributions (RDs)
of stars with iron abundance higher (HIA) and lower (LIA) than the
mean sample value as a {\it diagnostic} of a radial trend or segregation.
The radial segregation between the sub-samples of LIA and HIA RGs,
in the sense of the former stars being more concentrated to the
centers of their parent GCs, was found using a Kolmogorov-Smirnov
(K-S) test to be fairly high (P = 92\%) in NGC 6809 (M55) and
statistically significant in 47 Tuc, NGC 1851, NGC 3201, and NGC 6752.
Interestingly, three of the four later GCs are highly concentrated,
with central concentration $c > 2.00$ \citep{kravtsov13}.

As for NGC 3201, \citet{simmereretal13} determined iron abundance
for a sample of 25 its RGs [using 21 previously obtained archival
FLAMES-UVES spectra ($R\sim40,000$) and 5 of their MIKE spectra 
($R\sim40,000$)], and confirmed a spread in iron content found 
previously by \citet{gonzawaller98}.
According to our estimate \citep{kravtsov13}, the difference between
the RDs of LIA and HIA RGs of this sample turned out to be at a
marginally statistically significant confidence level of P = 96\%,
despite a fairly limited sample size. Note that the spectra used by
\citet{simmereretal13} are of higher resolution, approximately twice
as much, than those used by \citet{carrettal09}.

Later, \citet{mucciaretal15} re-analyzed the same FLAMES-UVES
spectra and came to a different conclusion about iron abundance spread
in the GC. They found that while their estimates of the [FeI/H]
ratio are very similar to those published by \citet{simmereretal13},
the intrinsic spread of iron abundance derived from FeII lines, in
contrast, turned out to be much smaller. The authors explained the
larger dispersion in iron abundance estimates made by
\citet{simmereretal13} was due to a spurious effect, originating from
unaccounted NLTE effects in the analysis of possible AGB stars
misidentified as RGB stars. \citet{mucciaretal15} concluded that
NGC 3201 is a monometallic cluster, with no evidence of iron spread.

\begin{figure}
\centerline{
\includegraphics[clip=,angle=-90,width=16.0 cm,clip=]{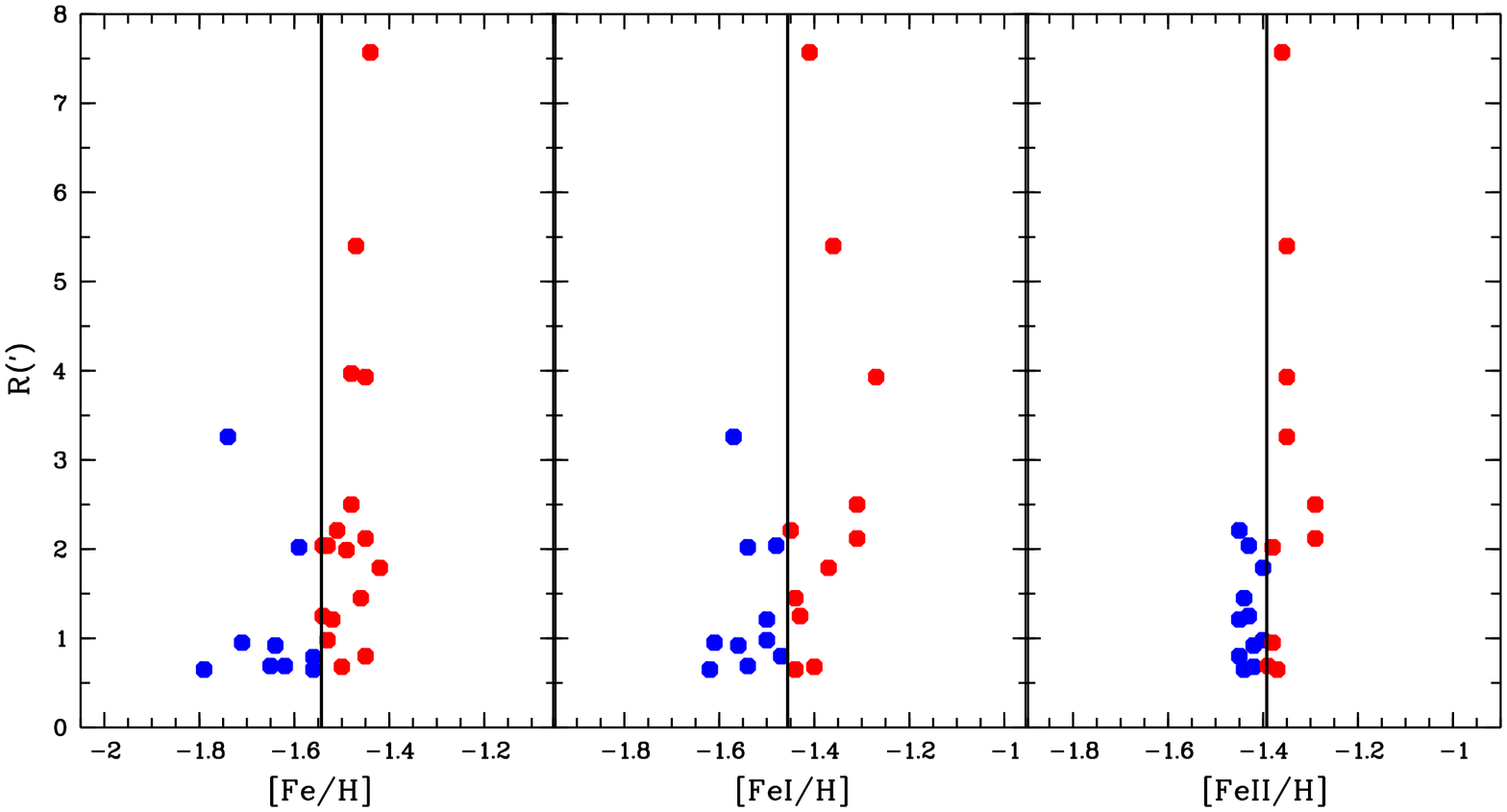}}
\caption{Dependence of iron abundance in RGs on their radial
distance from the center of NGC 3201, based on the data obtained by
\citet{simmereretal13} from FLAMES-UVES (21 stars) and MIKE (5
stars) spectra (left panel) and by \citet{mucciaretal15} from FeI
and FeII lines (middle and right panels, respectively) in the same
FLAMES-UVES spectra. The vertical continues lines mark the mean
values of [Fe/H].} \label{radtrend}
\end{figure}

Mucciarelli et al.'s (2015) conclusion is in disagreement with the
results of \citet{gonzawaller98}, who determined elemental abundances
in 18 RGs (using spectra with $R$ varying between 13,500 and 42,000) 
and estimated reddening for the stars. \citet{gonzawaller98}
found an iron abundance range of $\Delta$[Fe/H] $\sim 0.4 - 0.5$ dex,
four or 5 times as large the uncertainties in iron abundance
derived for an individual RG. They argued that this spread cannot be
attributed to the reddening variation and speculated that it could be
caused by a some kind of systematic effect ``resulting from an
unmodeled phenomenon in the atmospheres of cool giants, such as
chromospheric heating and back-warming...". We refer the reader to
this paper for more analysis of possible systematic effects. The
\citet{gonzawaller98} sample of RGs extends over a much larger
radial distance from the cluster center, closely approaching the
cluster's tidal limit which is $28\farcm45$ \citep{harris96}.
Then, the range in [Fe/H] found by \citet{gonzawaller98} becomes
more intriguing if it is related to how the [Fe/H] ratio depends
on the radial distance from the cluster
center (see discussion in Sect.~3 and Sect.~4). On the other hand,
\citet{mungeisvil13}, using high-resolution spectroscopy, found
significantly smaller dispersion of the [Fe/H] ratios derived for a
sample of 8 RGs, within $\Delta$[Fe/H] $\sim$ 0.12 dex, that could
be attributed to a real iron spread. However, their sample size is
much smaller as compared to \citet{gonzawaller98},
\citet{simmereretal13}, and \citet{carrettal09}. The furthest radial
distance from the cluster center reached by the sample stars in
\citet{mungeisvil13} is about 3.6 times as short as the cluster's
tidal radius, but it is comparable to those reached by other data,
except for the data of \cite{gonzawaller98}. The radial extent of
each spectroscopic survey analyzed in the present paper can be seen
in Fig.~\ref{radtrend} and Fig.~\ref{composite}.

In light of the result achieved by \citet{mucciaretal15} we first
study whether their improved estimates of iron abundance in NGC 3201
are free of the radial trend appropriate to other data sets.
Additionally, we include the \citet{gonzawaller98} data in analyzing
the dependence of iron abundance on projected radial distances (PRAD)
from the cluster center. In addition to the radial trend (of the
[Fe/H] ratio) under study, we show the presence of a notable same-sign
radial trend of oxygen abundance. Finally we discuss various effects
that could mimic the radial [Fe/H] trend observed in RGs.

\section{Data analysis}

The recently published archival data on iron abundance for 21 RGs in
the GC NGC 3201, obtained relying on both FeI and FeII lines in high
resolution FLAMES-UVES spectra and published by
\citet{mucciaretal15}, were taken from Table 1 of the paper. We
refer to this paper for more details about the spectra used, data
reduction and analysis, and the main conclusions reached. Moreover,
for the purpose of comparison, we used the data of
\cite{gonzawaller98} (their Table 5), \citet{carrettal09}, and
\citet{simmereretal13} (their Table 1). From the data of
\citet{mucciaretal15} on the iron abundance of individual sample
stars, we first calculated the mean values of the [Fe/H] ([FeI/H] or
[FeII/H], depending on the data used) ratio. Then, like in
\citet{kravtsov13}, we divided the total sample of RGs, N$_{tot}$,
into two sub-samples of RGs with iron abundance lower (N$_{LIA}$)
and higher (N$_{HIA}$) than its mean value, which were referred to
as LIA and HIA RGs, respectively. For the RGs of the sample, we
calculated both their rectangular coordinates X
($\Delta\alpha\arcsec$) and Y ($\Delta\delta\arcsec$) and PRAD
($R = \sqrt{X^2+Y^2}$), expressed in arcseconds, relative to the
cluster center, using the equatorial coordinates of the stars taken
from Table 1 of \citet{mucciaretal15}. In turn, the co-ordinates of
the cluster
center was taken from \citet{harris96}. Alternatively, the PRADs
plotted in Fig.~\ref{radtrend} and expressed in arcminutes were
taken from \citet{simmereretal13}. Finally, the probability
(expressed in percent) P that the two sub-samples have been drawn
from different RDs in NGC 3201 was estimated by applying a K-S test.

\cite{gonzawaller98} identified their sample stars with those from
the list of \citet{cotetal94}. We took the values of the PRADs of
the respective stars from this list and used them in our analysis
of the data of \cite{gonzawaller98}.

\section{Results}
\label{res}

\begin{figure}
\centerline{
\includegraphics[clip=,angle=-90,width=12.0 cm,clip=]{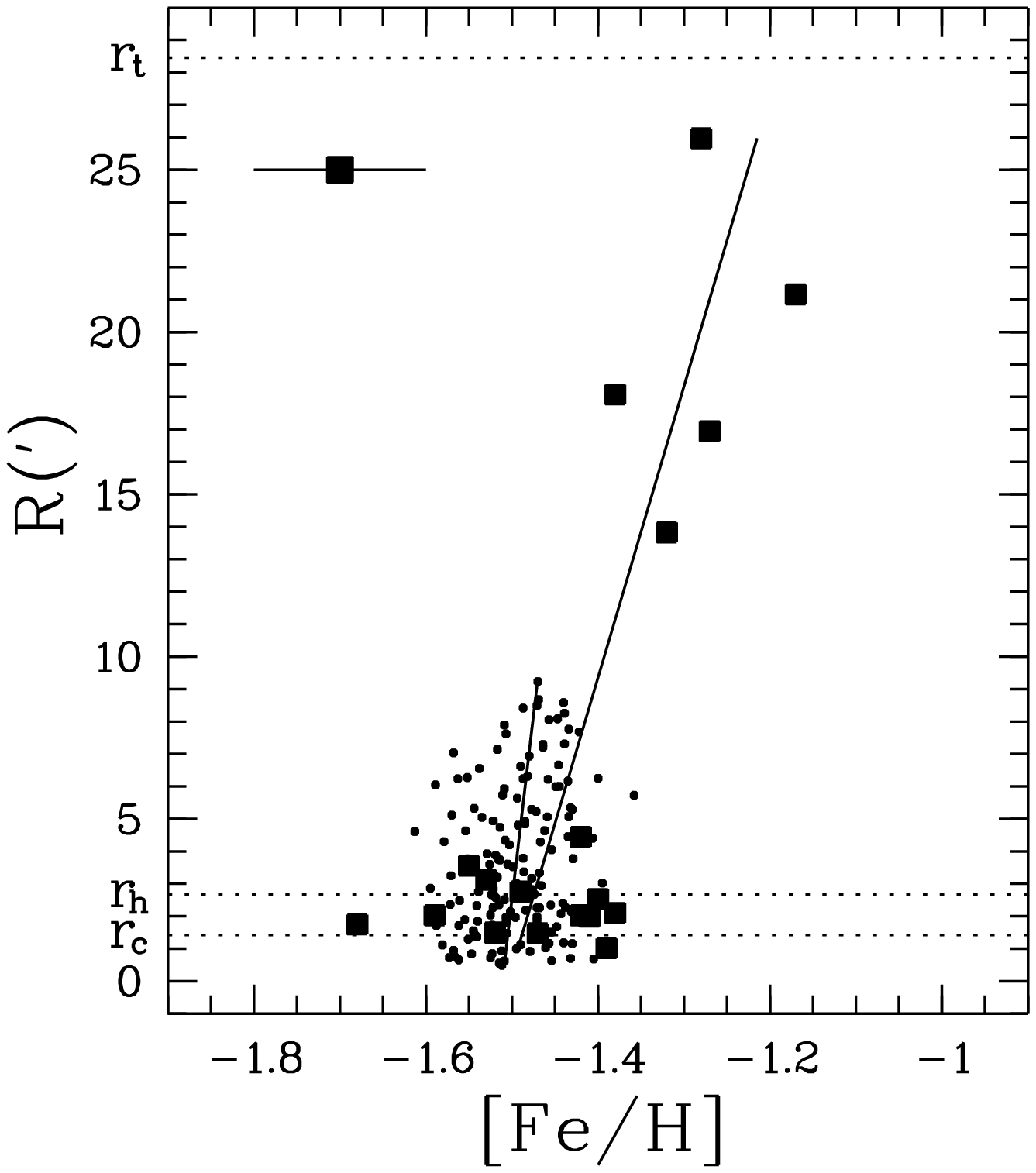}}
\caption{Comparison of the dependence of iron abundance in RGs on
their radial distance from the center of NGC 3201, inferred from the
data of \citet{carrettal09} (149 stars; dots) and
\citet{gonzawaller98} (18 stars; filled squires). The continuous
lines are linear fits to the data. The error bar in the upper left
corner of the figure shows the probable error estimated by
\citet{gonzawaller98} for the iron abundances they derived for
separate stars. The dotted horizontal lines mark the core, half-masse,
and tidal radii ($r_c$, $r_h$, and $r_t$, respectively) of NGC 3201.} \label{composite}
\end{figure}

\begin{figure}
\centerline{
\includegraphics[clip=,angle=-90,width=4.9 cm,clip=]{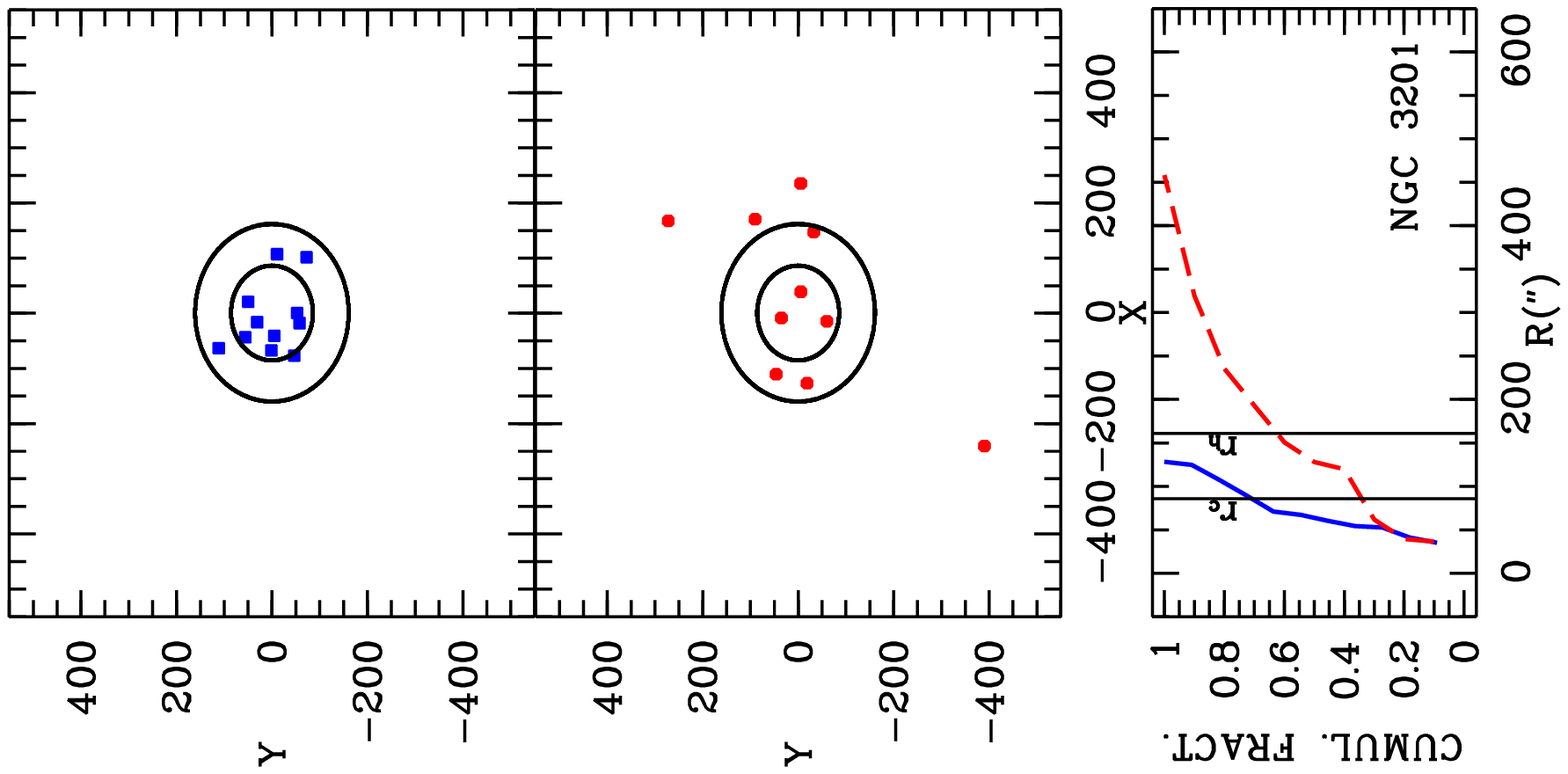}
\includegraphics[clip=,angle=-90,width=3.97 cm,clip=]{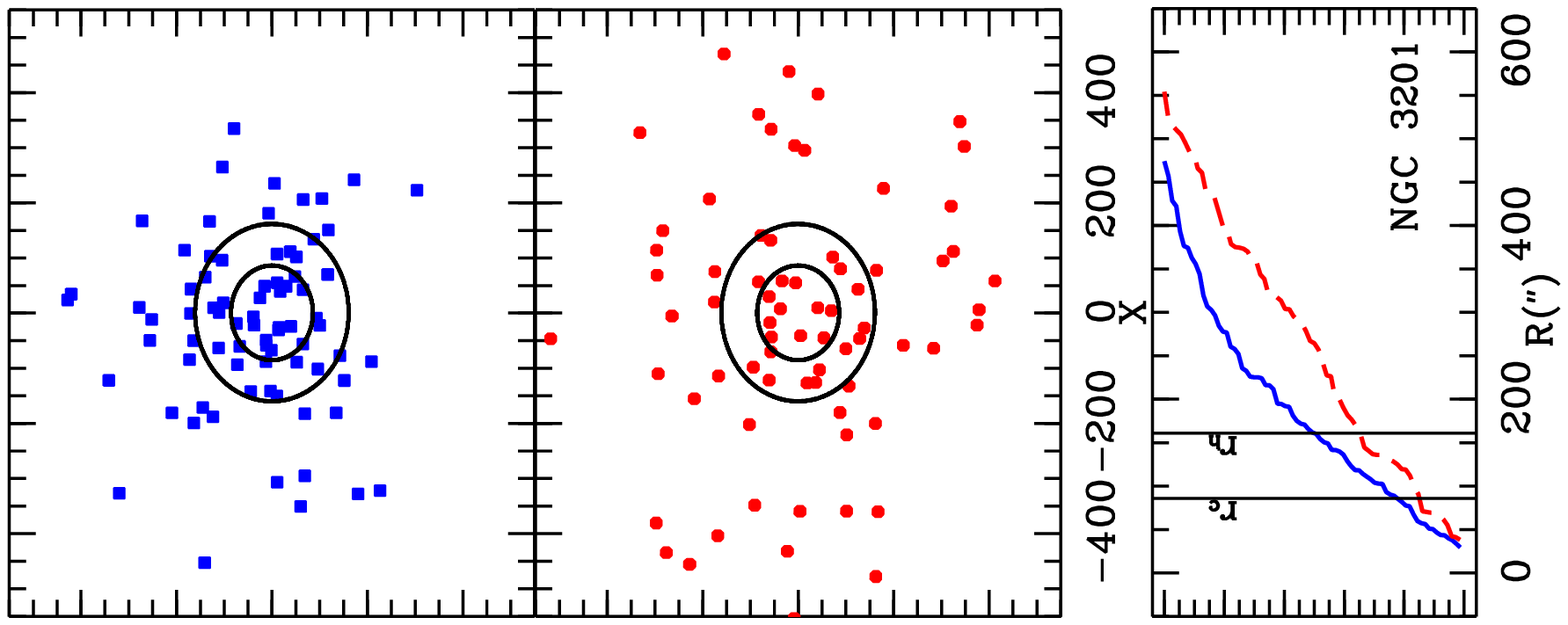}}
\caption{Positions (dots) and cumulative RDs (lines) of LIA (blue,
continuous) and HIA (red, dashed) RGs in the field of NGC 3201 are
compared for the data of \citet{mucciaretal15} and
\citet{carrettal09} in left and right columns of the panels,
respectively. Both the X,Y coordinates and PRAD, $R$, are in
arcseconds.} \label{posinfield}
\end{figure}

In three panels of Fig.~\ref{radtrend} we plot and compare the RDs
of iron abundance in the GC NGC 3201 as inferred from the following
sets of data: that of \citet{simmereretal13} shown in left panel
followed by data from \citet{mucciaretal15} for the [FeI/H] and
[FeII/H] ratio (in middle and right panels, respectively). As we
noted above, 21 stars are in common in both studies. The same plot
for the data of \citet{carrettal09} can be found in
\citet{kravtsov13} and below, in Fig.~\ref{composite}, too. We show
that the dispersion of points around the mean values of [Fe/H]
(marked by vertical lines in the panels) is comparable in the left
and middle panels, while it is indeed notably smaller in right panel
showing the RD of the [FeII/H] ratio. However, the most important
point for our study is to compare the RDs of LIA and HIA RGs for the
new data on iron abundance. For both [FeI/H] and [FeII/H] arrays of
the estimates, the LIA RGs again turned out to be notably more
centrally concentrated in the GC NGC 3201 than their HIA
counterparts. Quantitative estimates, using a Kolmogorov-Smirnov
test, support the apparent difference: despite a fairly limited
total sample size of 21 stars, the RDs of the two sub-samples of the
RGs are different at high confidence level of 92\%. Moreover, in
Fig.~\ref{posinfield} we show the positions and respective cumulative
RDs of the LIA and HIA RGs of the sample in the field of NGC 3201
and compare them with the same plots for the data of
\citet{carrettal09}. From this comparison, one can see that the
difference between the cumulative RDs of LIA and HIA RGs in the data
of Mucciarelli et al. is formally even larger than in the data of
Carretta et al. However, the sample size of the former data is
unfortunately significantly smaller, what has resulted in lower
confidence level.

For the least dispersed [FeII/H] ratio, we estimate the difference
and its uncertainty between the mean iron abundance in the central
and outer parts of the cluster. By drawing a boundary between the
two parts around the PRAD of $R = 2\arcmin$, we find the difference
is close to $\Delta$[FeII/H] = 0.05 $\pm0.02$ dex. In particular,
$\Delta$[FeII/H] = 0.054 $\pm0.022$ and $\Delta$[FeII/H] = 0.052
$\pm0.020$ dex for the boundary drawn at exactly $R =120\arcsec$ and
$R =107\arcsec$, respectively. It is evident that this is a very
approximate estimate of both $\Delta$[FeII/H] and $R$, due to the
limited sample size. The uncertainty is standard error of the
difference between the means of the two sub-samples and it has been
calculated from the errors of the mean ($\sigma_i/\sqrt{N_i-1}$) of
each sub-sample. In addition to the notable difference in
the RDs between the sub-samples of LIA and HIA RGs in NGC 3201, the
trend is also expressed in other quantitative form: mean iron
abundance in the central part of the cluster is lower, by 0.05 dex,
than in the outer one and it is 2.5 times as much as its error. The
estimated $\Delta$[FeII/H] is formally indistinguishable, within the
error, from our estimate of the magnitude of the trend (shown in
Fig.~\ref{composite}) originally made relying on [FeI/H] ratios
derived by \citet{carrettal09} for a much larger number of the
cluster RGs. We note that the standard deviations in both data sets,
those obtained by \citet{mucciaretal15} and by \citet{carrettal09},
are indistinguishable from each other and equal to $\sigma = 0.05$
dex. The error of the mean, however, is 2.6 times smaller in the latter
data set, since the sample size is 7 times as large as the sample
size of the former data. Thus, acknowledging that the fact that the
number of FeII spectral lines is fewer than the number of FeI
lines in RG spectra, the final scatter (expressed in standard
deviation of iron abundance estimates) reported by
\citet[using FeII lines]{mucciaretal15} is not smaller than the
scatter of the data derived by \citet{carrettal09} relying on FeI
lines.

The data of \citet{carrettal09} are plotted, in the $R$ vs [Fe/H]
diagram, with dots in Fig.~\ref{composite} and over-plotted with a
continuous line that is a linear fit to these. The same is shown for
the data on the iron abundance derived by \citet{gonzawaller98} for
their sample of 18 RGs. This figure shows that (1) the results on
iron abundance obtained in both studies for stars falling in the
same range of PRAD (i.e. at $R <$ 5\arcmin) from the cluster center
are in very good agreement, within the uncertainty, but (2) the data
of \citet{gonzawaller98} reveal much more pronounced radial trend of
iron abundance, since the outermost five
RGs (located at PRAD $R >$ 13\arcmin) have, on average, obviously
higher [Fe/H] ratio than the rest of the stars, 13 RGs, located
within $R \approx$ 5\arcmin. It is the essential difference in iron
abundance between these two groups of stars that increased the range
spanned by the [Fe/H] ratio derived by \citet{gonzawaller98}. The
scatter in each group, in isolation from one another, is obviously
lower than in the total sample. Indeed, the [Fe/H] ratio in the
outermost stars varies in the range of -1.38 $\leq$ [Fe/H] $\leq$
-1.17 dex, whereas the ratio in the inner RGs varies between -1.68
$\leq$ [Fe/H] $\leq$ -1.38 dex. The mean values of [Fe/H] and
standard errors of the means for the outer and inner groups are
[Fe/H]$_{out} = -1.28\pm0.04$ ($\sigma = 0.08$; 5 stars) and
[Fe/H]$_{inn} = -1.48\pm0.03$ ($\sigma = 0.09$; 13 stars) dex,
respectively. Hence the difference between the means and its
uncertainty are $\Delta$[Fe/H] = 0.20 $\pm0.05$ dex, in case of no
systematic error. Therefore, the value of $\Delta$[Fe/H] is 4 times
as large as its formal random error that is standard error of the
difference between the means of the two sub-samples. However, as was
noted above, \citet{gonzawaller98} suppose the metallicities they
derived may be affected by a systematic error. It is implied by
probable dependence of the estimated iron abundances in the sample
RGs on the temperature of the stars. It is easy to see that the
outer RGs have, on average, higher temperatures. So, the metallicity
of the five outer RGs may really be lower and, therefore, the magnitude
of the radial trend may be smaller than 0.2 dex. It is apparently
supported by the result of \citet{coveyetal03} who re-investigated
the metallicity spread in NGC 3201 reported by \citet{gonzawaller98}.
For the purpose of their study,
\citet{coveyetal03} gathered new spectra for six of the eighteen RGs
studied by \citet{gonzawaller98}. The rederived iron abundances for
these six RGs, using spectroscopic temperatures, resulted in
somewhat smaller metallicity difference between the most and least
metal rich stars (C\^{o}t\'{e} 8 and C\^{o}t\'{e} 246, respectively,
from the list of \citet{gonzawaller98}), but at the same mean error
as compared to the results of \citet{gonzawaller98}. It should be
noted, however, that C\^{o}t\'{e} 8 is the only of the five most
metal-rich and most distant RGs in the sample of
\citet{gonzawaller98}. Moreover, C\^{o}t\'{e} 8 was not used to
rederive iron abundance by relying alternatively on photometric
temperatures. Overall, \citet{coveyetal03} find an intriguing offset
between metallicities they derived from photometrically and
spectroscopically determined parameters.

The \citet{coveyetal03} study could not quantify the difference in
iron abundance between the outer RG population and those that are
more centrally concentrated. Our present analysis shows, therefore,
that a reliable answer to this question would have important
implication for the formation and evolution of the cluster stellar
populations.

Fig. 3 of \citet{mungeisvil13} shows the CMD of NGC 3201 with their
data set, that of \citet{carrettal09}, and \citet{simmereretal13}.
The \cite{gonzawaller98} data set is shown in Fig. 2 of their paper.

\section{Discussion}
\label{discus}

We find that the data on iron abundance in the RGs of NGC 3201,
analyzed in our original study \citep{kravtsov13} and in the present
work, show increasing iron abundance with increasing radial distance
from the cluster center, irrespective of both the facility used to
gather observations and the methods of data reduction applied. All
but one of the previous studies are based on observations gathered
with fiber-fed multi-object spectrographs, except for five stars in
the sample studied by \citet{simmereretal13}, which were separately
observed with the MIKE spectrograph, presumably in standard slit
observing mode. To our understanding, the only total set of spectra
used by \citet{gonzawaller98} to derive their data, were obtained in
standard slit observing mode. Therefore, it is of low probability
that the trend of the same sense (iron abundance decreasing toward
the center of NGC 3201) is reproduced with different data by chance.

\begin{figure}
\centerline{
\includegraphics[clip=,angle=-90,width=12.0 cm,clip=]{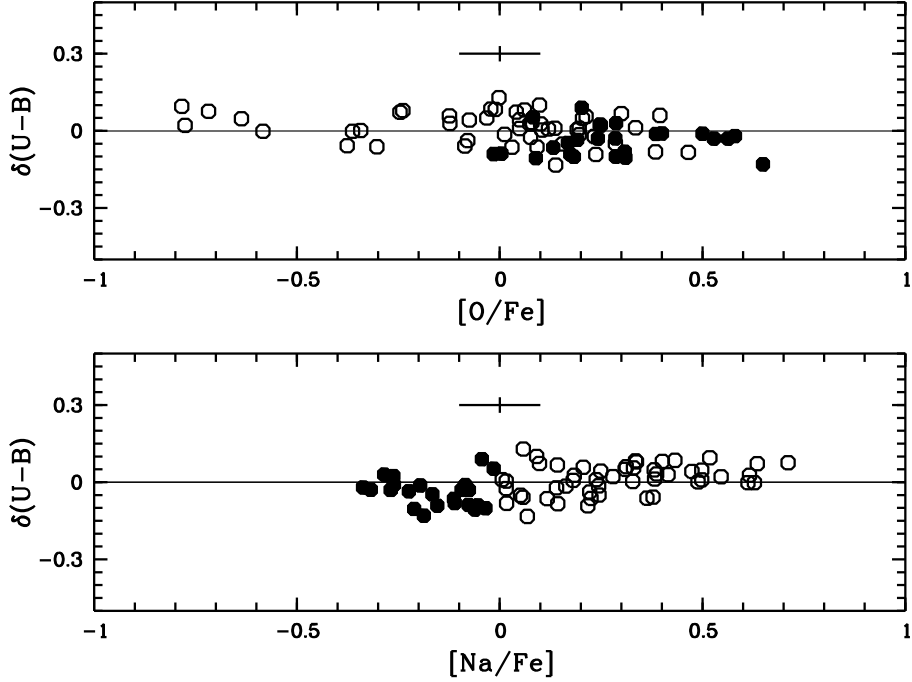}}
\caption{Dependence of of the $(U-B)$ color of RGB stars, expressed as their
deviations $\delta(U-B)$ from the RGB ridgeline in dereddened $U-(U-B)$ diagram
\citep{kravtsovetal10}, on the O and Na abundances (in the upper and lower
panels, respectively) derived by \citet{carrettal09} for the same stars.
Filled circles show RGs defined by the latter authors as first-generation
stars in NGC 3201, i.e. with [Na/Fe]$< 0$. The horizontal and vertical error bars
show typical uncertainties in the abundance estimates and in the $(U-B)$ color,
respectively. The latter is formally that of the photometry and it does not
include the uncertainty in dereddening corrections applied.} \label{abundUBphotom}
\end{figure}

The decrease in iron abundance towards the core of NGC 3201 might be
real. If real, the primordial (first generation) population of the GC
should be concentrated in the core of NGC 3201, as found by by
\citet{larsenetal15} in the GC M15. In the case of NGC 3201,
however, such a possibility seems to be improbable, because there is
indirect evidence that the nitrogen enriched population of RGB stars
tends to be more concentrated to the center of the cluster. This is
implied by the result of \citet{carrettal10} who coupled Na (and O)
abundances for a limited sample of the cluster RGB stars with
multi-color photometry available for a much larger number of RGB
stars \citep{kravtsovetal09}, for which \citet{kravtsovetal10} showed
a correlation between $U$-based photometric characteristics and
radial distance from the cluster center. In
Fig.~\ref{abundUBphotom}, we summarize a relationship between
the sodium and oxygen abundances derived by \citet{carrettal09} for
RGB stars in NGC 3201 and their above-mentioned photometric
characteristics as dependence of deviations $\delta(U-B)$ of these RGs
from ridgeline of the RGB, obtained by \citet{kravtsovetal10}
from the cluster $U-(U-B)$ diagram corrected for differential reddening
(see the referenced paper for more details). We showed by filled circles
the RGB stars defined by \citet{carrettal09} as first-generation stars.
An additional point concerning this dependence is that the deviation
$\delta(U-B)$ of RGs at a given abundance (of either sodium or oxygen)
is, on average, larger at larger PRAD from the cluster center. This
supplementary relation increases the width of the dependence along the
$\delta(U-B)$ axis and makes it larger than the typical uncertainty
of the photometry. Or in other word, at a given $\delta(U-B)$ RGs with
higher oxygen (lower nitrogen) abundance are, on average, at a larger
PRAD from the cluster center. Iron abundance does not show any
detectable systematic variation between the extreme values of
$\delta(U-B)$. The mean [Fe/H] value is within the error of the
mean at any $\delta(U-B)$.

Although being of low probability, a somewhat decreased iron abundance
in the central part of NGC 3201 might be due to presently unknown
detail(s)of star formation in GCs and subsequent dynamical
evolution of the hosts. Our knowledge about the formation of
massive (globular) star clusters is poor so far. There is a
permanent inflow of new, sometimes challenging data on (radially
depending relationship between) the kinematical and chemical
characteristics of stellar populations in GCs. In particular,
for a sample of the turnoff stars of 47 Tuc,
\citet{kucinskasetal14} find the radial dependence of oxygen
abundance and, moreover, a significant correlation between
this abundance and the velocity dispersion of the stars. Also,
\citet{fabriciusetal14} find evidence of surprisingly ubiquitous
central rotation (and its tight correlation with the central
velocity dispersion) in a sample of eleven GCs, which can be
assumed to be related to kinematically decoupled different
stellar populations in the GCs.

It is obvious that the spectroscopic data is for the brighter RGs,
with luminosity exceeding that of the RGB bump. The only exception
is the sample of RGs studied by \citet{carrettal09}, which includes
a large number of stars below the RGB bump. In general, there is no
guarantee that the data on elemental abundances in brighter RGs will
agree with those in lower RGs and particularly in main sequence
stars \citep[e.g.,][and references therein]{grattonetal00}.

One could suppose that a systematic error originated from photometric
factors due to erroneous corrections applied for differential
reddening. A systematic effect might arise from the same type of
multi-object fiber-fed spectroscopy used to gather the bulk of
observations. The sample of stars located at different radial
distance from the cluster center and might be affected by a
systematically increasing sky background towards the center. The
crucial condition for the varying sky background to affect the
estimates of iron abundance in the central part of a GC seems to be
a radially varying spectral composition of the background (i.e.,
varying its color) rather than increasing intensity at a constant
color. Any notable color gradient of the background light, an
important contributor to the light of any GC, should result in the
radial color variations of the GC itself. In the last three decades,
such variations were found in GCs, in the sense that the clusters were
judged to be bluer, in addition to the strengthening of Balmer
absorption lines, towards the centers of the GCs. This effect was
discussed in a review of the topic by
\citet[and references therein]{djorpiot93}. It is worth mentioning
that most of these GCs are post-core-collapse objects and the gradient
typically extends to at least a few tenths of arcsec in radius and
even up to $\sim100\arcsec$ or further. ``The typical $(B-I)$ gradient
amplitudes are $\sim0.1-0.3$ mag per decade in radius", according to
estimates by \citet{djoretal91}. The weakest point of this argument
is that it is unknown whether or not NGC 3201, which is irregularly
reddened across its face, has a color gradient towards its center.
This cluster is not a post-core-collapse system. The reality of the
gradient of dereddened color is subject to dispute. An additional
problem is to quantify the effect (if any) of the color gradient on
iron abundance estimates deduced from a multi-object fiber-fed
spectroscopy.

\begin{figure}
\centerline{
\includegraphics[clip=,angle=-90,width=8.2 cm,clip=]{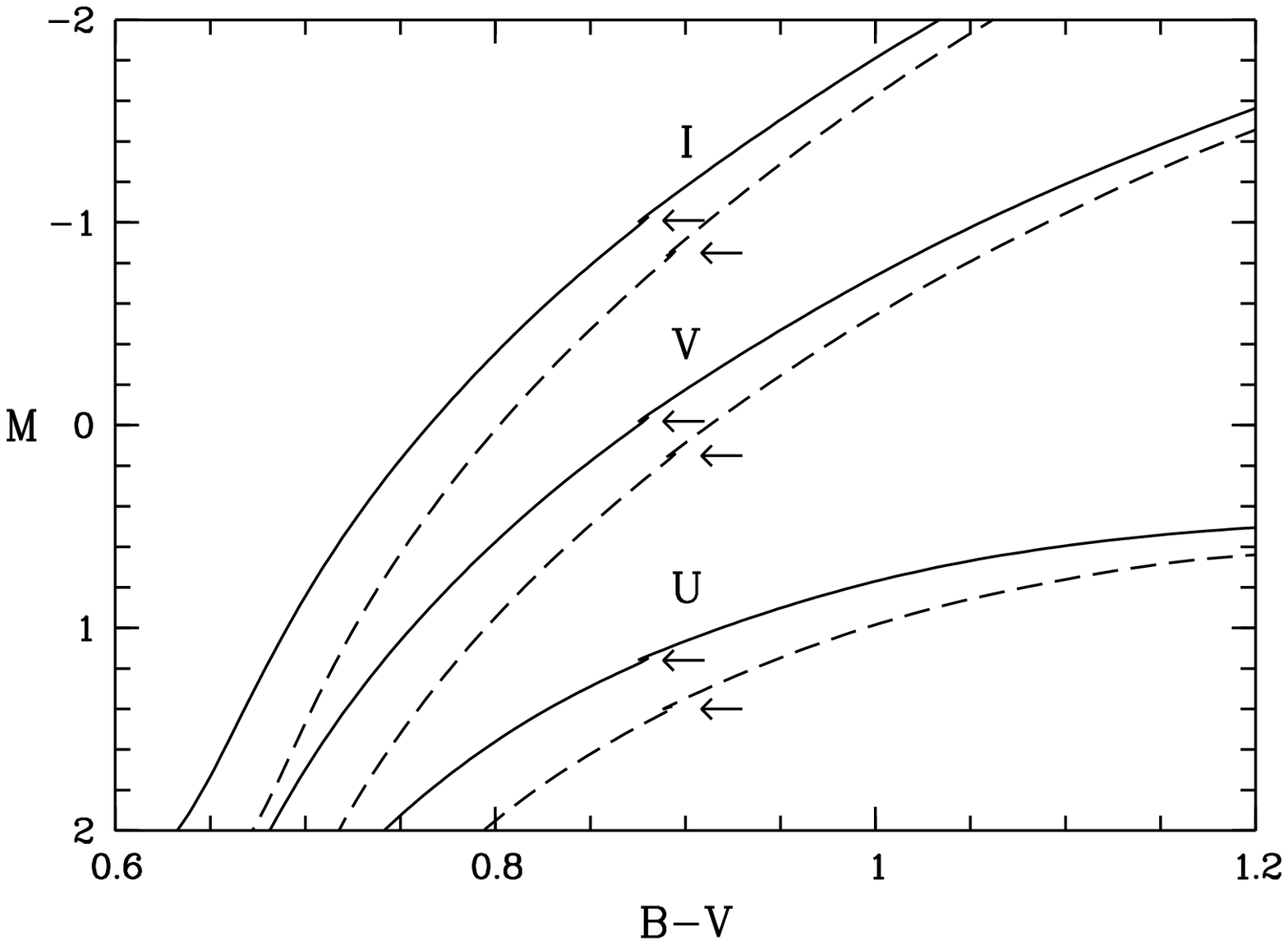}}
\caption{Two evolutionary tracks \citep{valcarcetal12} are shown
in CMD with the $(B-V)$ color and different absolute magnitudes
in the $I$, $V$, and $U$ passbands. The continuous and long-dashed
lines are the tracks computed for the same age, iron and helium
abundances, [Fe/H]$=-1.40$ and Y=0.250, but for different mass and
overall metallicity: $M_1 = 0.80 M_{\sun}$; Z$_1$=0.00069 ([O/Fe]=0.0)
and $M_2 = 0.81 M_{\sun}$; Z$_2$=0.00115 ([O/Fe]=0.3), respectively.
The arrows indicate the predicted position of the peculiarity in the
models' RGB evolution, which is responsible for the occurrence of the
RGB bump.} \label{bumpintracks}
\end{figure}

One might suggest a radial trend of the mass of coeval RGB stars at
the same evolutionary stage, as another alternative for mimicking
the iron abundance trend in NGC 3201. Such a possibility seems not
to be as exotic as could be at first
glance. On the contrary, the opposite assumption of no mass
variation would rather be improbable, given observational evidence
of significant variation of the abundance of key chemical elements
in stars of several GCs. In particular, nitrogen abundance, [N/Fe],
is known to span about two orders of magnitude in RGB stars in NGC
6752 \citep{yongetal08}. In a general case, such large abundance
variations in a GC, even being significantly compensated by the
anti-correlated abundances of other main contributors (i.e., the
oxygen) to overall metallicity, can result in overall metallicity
variations (at a constant iron abundance) of the order of a few
tenths of dex in the GC stars. The present-day GC stars of the same
age, but varying metallicity, might have different masses. In the
case of initial radial segregation of stars of different
metallicity in a GC, one might expect that more centrally
concentrated metal-rich stars would conserve their preferable
central location in the GC, during a Hubble time, due to their
increased mass as compared to the counterparts of lower metallicity.
How could the mass difference be due to varying overall
metallicity at a given constant iron abundance? We estimated it for
this simplest case by computing two evolutionary tracks
\citep{valcarcetal12} using two models of the same iron and helium
abundances, [Fe/H]$=-1.40$ and Y$=0.250$, but of different mass and
overall metallicity which allow the two tracks to achieve the end of
RGB evolution at the same age. In more details: (1) we imposed the
values of the initial mass, $M_1$, and the overall metallicity,
Z$_1$, for one model and computed evolutionary track between the
zero-age main sequence and the RGB tip as well as the duration of
the track; (2) then we imposed a higher metallicity, Z$_2$ (Z$_1 +
\Delta$Z), and through iterative computations of models with this
metallicity and varying mass, we searched such $M_2$ that its
combination with Z$_2$ resulted in an evolutionary track with the
same total lifetime as that of the track computed for the combination of
Z$_1$ with $M_1$. By this method, the effect of varying metallicity on
age was compensated by varying mass {\it at the same iron
abundance}, [Fe/H]$=-1.40$, corresponding to NGC 3201. We finally
estimated $M_2$ corresponding to the following combinations of
varying mass with a plausible variation of metallicity in NGC 3201:
$M_1 = 0.80 M_{\sun}$; Z$_1=0.00069$ ([O/Fe]$_1=0.0$) and $M_2 =
0.81 M_{\sun}$; Z$_2$=0.00115 ([O/Fe]$_2=0.3$). We note that this
does not mean at all that [O/Fe] should just be higher at higher Z.
Instead, a higher [N/Fe] would be more natural. However,
unfortunately, there was the only possibility to vary Z at the
same [Fe/H] in the models computed. We used two available
options for oxygen abundance, differing by 0.30 dex. This makes the
mass difference as large as $\Delta M = 0.01 M_{\sun}$. This is too
small for a radial segregation of stars due to the dynamical
evolution of a GC with no initial radial mass segregation
\citep[see more details in][]{larsenetal15}. For this mass
difference and at the metallicity of NGC 3201, the temperature
difference between the tracks ($\Delta$$T_\mathrm{eff} \approx
40~\mathrm{K}$ corresponding to $\Delta(B-V)\approx$ 0.03 mag), in
particular at the level of the RGB bump or above it, is comparable
(in modulus) with the value necessary to result in the error of iron
abundance of order $\Delta$[Fe/H]$\sim$0.05 dex. Note that,
according to \citet{natafetal11}, the same difference in [Fe/H]
corresponds to $\Delta$$T_\mathrm{eff} \approx 17~\mathrm{K}$ at the
metallicity of 47 Tuc (i.e. [Fe/H]$=-0.75$). The more important
point is that in the framework of this simplified scenario for the
considered models they predict radial dependence of photometric
effects, some of which are in apparent disagreement with the observed
ones. Specifically, the models predict decreasing brightness of the
RGB bump toward the cluster center by magnitude depending on the
photometric passband. In particular, the variation of the bump level
is expected to be around $\Delta V \approx$ 0.18 mag, and to be larger
and slightly smaller in $U$ and $I$, respectively
(Fig.~\ref{bumpintracks}). The expected dependencies of the the RGB
bump magnitude on both radial distance from the cluster center and used
passband are in disagreement with the apparent behavior of the bump
level in NGC 3201 as implied by cluster photometry obtained by
\citet{kravtsovetal09}.

No significant radial trends are normally expected in GCs, given
their typical relaxation time. In general case, therefore, a radial
variation of the RGB bump level (in the $V$ magnitude) in a typical
GC is expected to be none or negligible, i.e. within the threshold
of its real detectability. In any case, reliable detection of such a
variation of the RGB bump brightness of the order of $\Delta V \sim$
0.05 mag is a challenging task. It can have credibility provided
that, besides known requirements imposed on both the accuracy of
photometry and sample size of RGs belonging to the RGB bump, there
must be insignificant uncertainty in the reddening variation across the
cluster face, within $\delta E(B-V)<$0.005 mag, so that the
uncertainty in extinction variation would not be more than the
typical r.m.s. of photometry, $\delta A_\mathrm{v}\sim\sigma\sim
0.010 - 0.015$ mag. Unfortunately, NGC 3201 does not meet
this condition, since its reddening is highly and irregularly
variable \citep[e.g.,][]{vbramat01}. Any reddening-corrected photometry
would have $\pm$0.05 mag. measurements with residual errors in
reddening correction of the order of $\delta E(B-V)\sim\pm$0.015,
i.e. three-four times as large as the above-mentioned limiting
value. This is converted in residual magnitude error $\delta
V\sim\pm0.05$ mag due to the uncertainty in reddening correction.
This can make it hardly possible to reliably establish and quantify
RGB bump variation within $\Delta V  < $ 0.10 mag in NGC 3201 as was
observed 47 Tuc \citep{natafetal11}, especially taking into account
a notably smaller number of RGs in the former GC.

\begin{figure}
\centerline{
\includegraphics[clip=,angle=-90,width=8.0 cm,clip=]{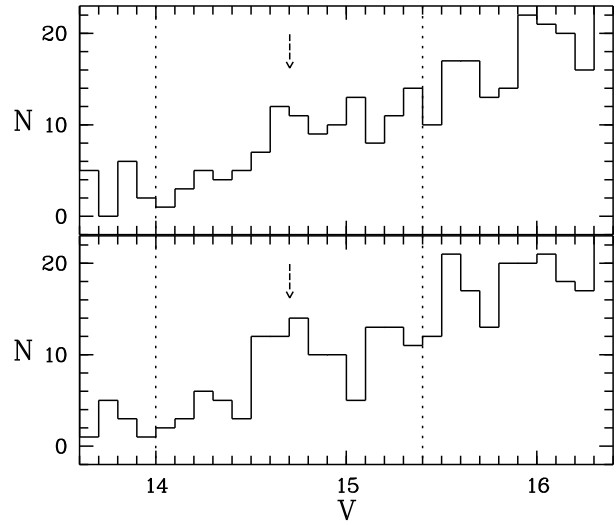}}
\caption{A part of the RGB LF obtained by \citet{kravtsovetal09} in
the $V$ magnitude corrected (top panel) and uncorrected (bottom
panel) for differential reddening in NGC 3201. The arrows mark the
position of the RGB bump, while the dotted lines show the
voluntarily imposed range of the bump region. Dependence of its LFs
in both the $V$ and $U$ magnitudes on PRAD from the cluster center
is demonstrated in Fig.~\ref{bumpraddep}.} \label{LFRGB}
\end{figure}

To draw any conclusion about this subject, we examined radial
behavior of the RGB bump luminosity function (LF) in three
passbands. In $V$ and $U$, we used photometric data corrected and
uncorrected for differential reddening and uncorrected in $I$. To
achieve this goal, we relied on multi-color photometry made by
\citet{kravtsovetal09} in a fairly large cluster field. In
particular, we used the same data sample on RGs which were selected
and decontaminated by the authors in their original analysis of the
cluster RGB LF (note, however, that the same result is achieved
without any decontamination). Thus, we refer the readers to this
paper for more details relevant to the subject. Fig.~\ref{LFRGB}
shows a fragment of RGB LFs in the $V$ magnitude corrected (top
panel) and uncorrected (bottom panel) for differential reddening in
NGC 3201. The arrows mark the resulting apparent position of the RGB
bump in the LFs for the total sample of RGs, whereas the dotted
lines show the $V (V_c)$ magnitude range of $\Delta V (V_c) = 1.4$
mag tentatively defined as RGB bump region, i.e. within $\pm0.70$
mag around RGB bump position.

\begin{figure}
\centerline{
\includegraphics[clip=,angle=-90,width=16.0 cm,clip=]{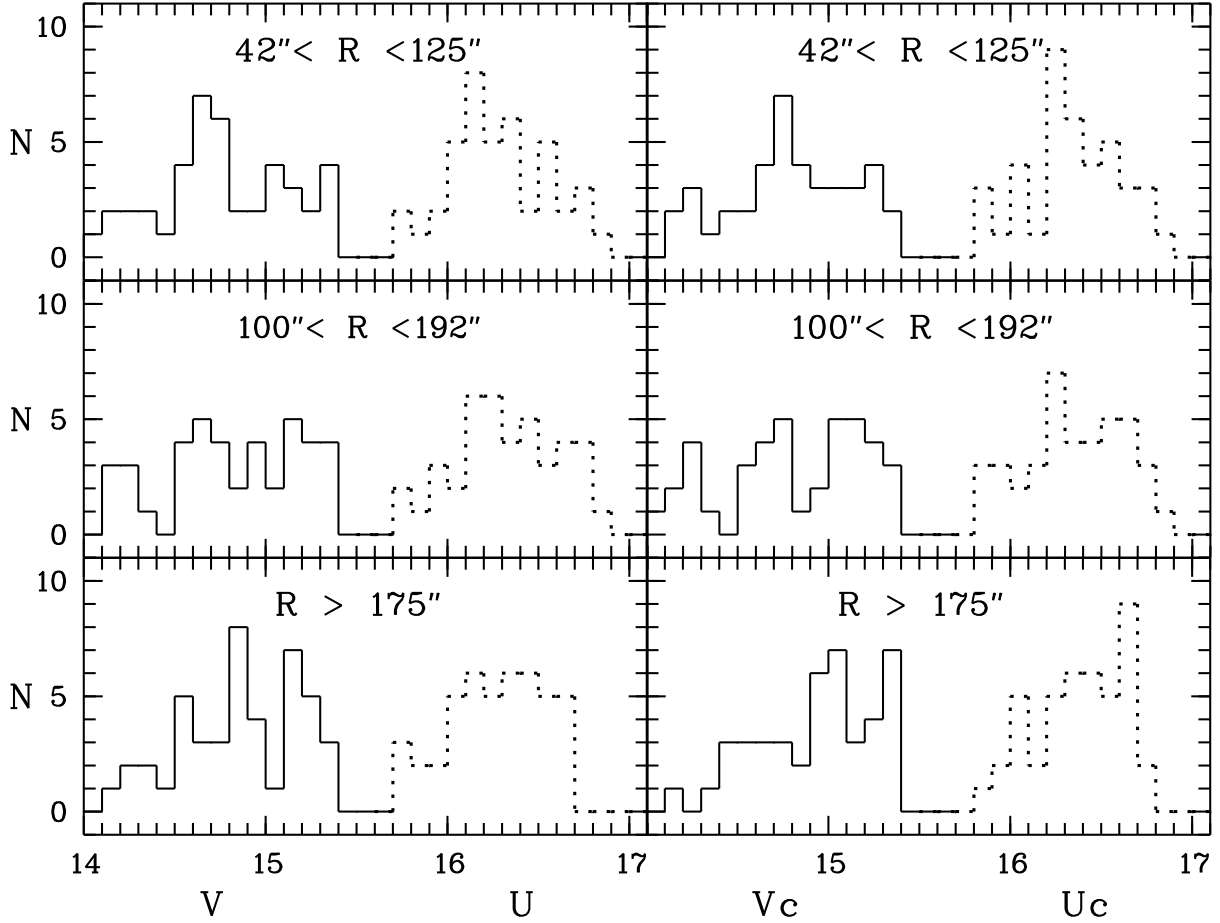}}
\caption{Dependence of the LF of the RG bump region, defined as
shown in Fig.~\ref{LFRGB}, in the $V$ (continuous line) and $U$
(dotted line) passbands on PRAD from the cluster center in a
14$\arcmin$x14$\arcmin$ cluster field. Panels in left and right
columns show the LFs based on red-giant magnitudes uncorrected ($V$;
$U$) and corrected ($V_c$; $U_c$) for differential reddening
\citep{kravtsovetal09}, respectively. Three panels, from top to
bottom, in each column show LFs for three consecutive cluster
regions with increasing PRAD from the cluster center. Indicated in
the panels are the ranges of the PRAD, $R$ (expressed in
arcseconds), limiting each region. The upper limit of the PRAD of
RGs in the outermost region is $R_{max}\approx10\arcmin$.}
\label{bumpraddep}
\end{figure}

As seen in Fig.~\ref{LFRGB}, we scan a broad RGB bump region to
show the structure of the bump, while noting that the number of RGs
in this region is limited. The latter
factor did not allow us to reliably estimate the position of
the bump at different PRAD from the cluster center. Thus, our main
goal was to examine whether there is a tendency in the radial
variation of the bump level and to make rough quantitative
estimates. The LFs in $U$ and $U_c$ (and in $I$, too) were obtained
simply using respective magnitudes of the RGs selected by their $V$
and $V_c$ magnitudes. In other words, the LFs of the RGB bump region
in $U$ and $U_c$ (and in $I$) are exactly of the same RGs that were
selected by their $V$ and $V_c$ magnitudes. In
Fig.~\ref{bumpraddep} we show dependence of the LF of the RGB bump
region on PRAD from the cluster center. These LFs are for three
overlapping cluster areas with increasing (from top to bottom panel)
mean PRAD from the cluster center. The number of RGs falling into the
specified bump range is virtually the same in each area. Since the
photometry used was made in a square area (14$\arcmin$x14$\arcmin$)
centered on the cluster, the external border of the outer region is
not circular and the longest PRAD of stars in this region,
$R_{max}\approx10\arcmin$, is achieved in the corners of the area.
To increase the reliability of our analysis, we excluded the
central cluster, within PRAD $R < 42\arcsec$ (100 pix), from
our consideration, thereby at least reducing potential problems that
might be caused by a systematic error due to crowding effects and
possible somewhat different completeness at fainter and brighter
magnitude of the bump just in the central cluster region. The
photometry by \citet{kravtsovetal09} was shown to be in very good
agreement with that by \citet{ste00} in $V, I$. The agreement with
photometry by \citet{laysar03} was also very good in the $V$ passband
but somewhat poorer in the $I$ passband due to a systematic
difference between the two data sets, that was magnitude-dependent.

\begin{figure}
\centerline{
\includegraphics[clip=,angle=-90,width=16.0 cm,clip=]{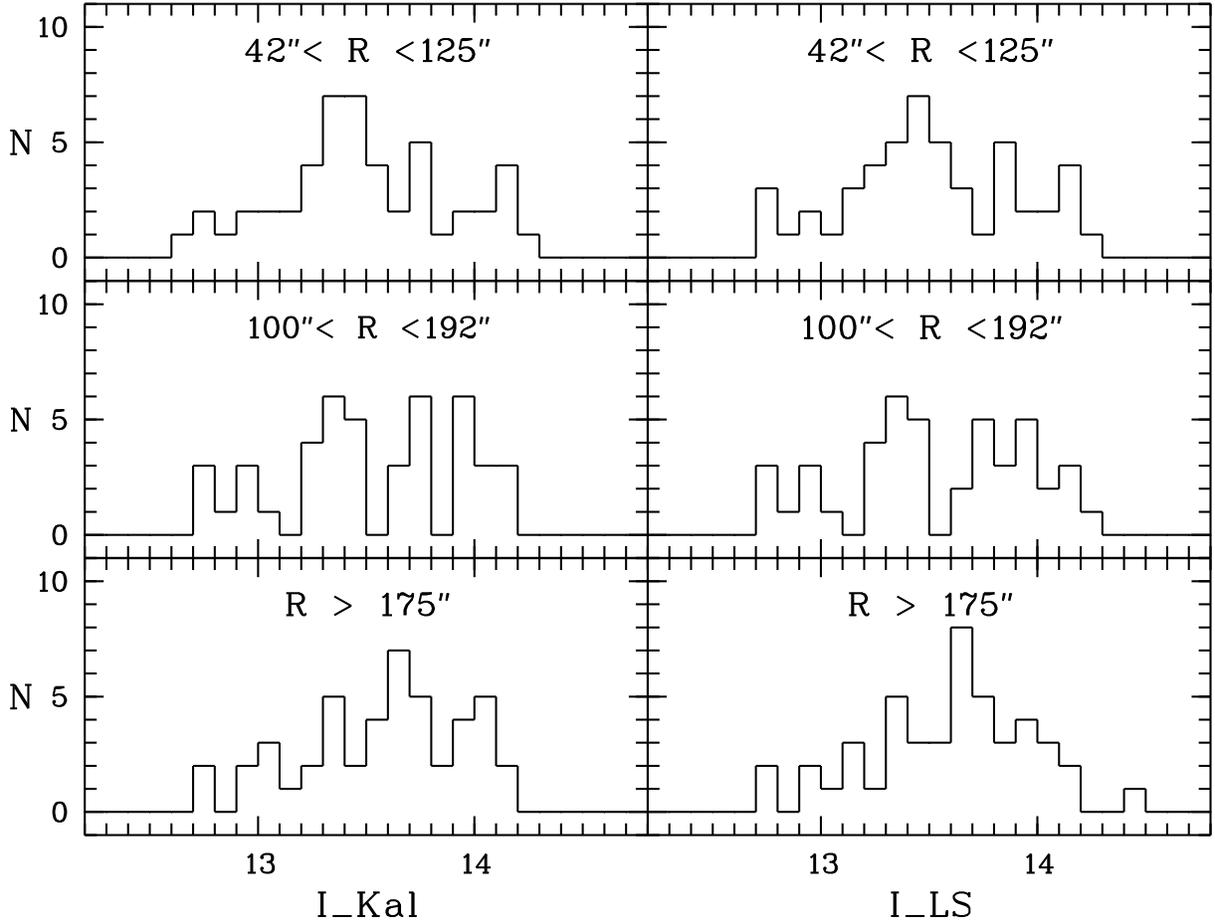}}
\caption{Comparison of dependence of the RG bump region LF,
defined as shown in Fig.~\ref{LFRGB}, in the $I$ passband on PRAD
from the cluster center. The left and right panels show the
dependence based on photometry of \citet{kravtsovetal09} and
\citet{laysar03}, $I_{Kal}$ and $I_{LS}$, respectively. The 
uncertainty in the number of stars, N, falling in each bin in all 
the histograms showing RGB bump LFs (including Fig.~\ref{bumpraddep}) 
is $\pm\sqrt{N}$.}
\label{bumpindifIbands}
\end{figure}

The LFs in the $V$ magnitude show a tendency of displacement of the
RGB bump level towards fainter magnitude in the outer part of NGC
3201; we detect an apparently decreasing proportion between
the number of RGs at $V\leq 14.70$ and that at fainter
magnitude. One can note also that (1) this tendency is the same, in
principle, in both $V$ and $V_c$ magnitudes and (2) a notable
transformations in the LFs begin to occur somewhere between PRAD
$2\arcmin < R < 3\arcmin$. In the central part of the cluster, the
bump is surely detected and confidently fixed at $V\approx 14.70$.
Estimating the exact position of the bump in the outer part, at $R >
175\arcsec$, is unreliable. The displacement of the bump towards
fainter magnitude may be as large as $\Delta V \sim0.15$ mag. Here,
we have to stress that this variation cannot be explained by a
spurious brightening of RGB stars towards the center of NGC 3201
due to a systematic effect of the photometry in the crowed central
region. This possibility is in fact ruled out, since
\citet{natafetal13} found the bump location at $V = 14.649 \pm0.032$
relying on $HST$ photometry of the (central part of the) GC NGC 3201.
This bump location is in very good agreement with and virtually
indistinguishable from that seen in the left right panel of
Fig.~\ref{bumpraddep}. The LFs in the $V_c$ magnitude show virtually
the same radial variation of the bump position. Relying on these LFs
the displacement can be estimated as being as large as
$\Delta V_c \sim 0.20$ mag. Separately, the left panes of
Fig.~\ref{bumpindifIbands} show
dependence of the RGB bump region LF in the $I$ passband on PRAD
from the cluster center. For comparison, the same dependence relying on
photometry made in the cluster by \citet{laysar03} is presented
in right panels. The sample stars are those in common in
the two datasets. One can note that (1) the radial behavior of the
RGB bump LF in the $I$ and $V_c (V)$ (Fig.~\ref{bumpraddep}) passbands
is very similar and (2) the LFs obtained from the two datasets show
virtually indistinguishable radial dependence in spite of some
disagreement between photometries by \citet{laysar03} and
\citet{kravtsovetal09} in the $I$ passband.

\citet{kravtsovetal09} estimated differential reddening in
the observed field of NGC 3201 from shifts of the CMDs, relative to
each other, of elemental areas composing the field ``provided reddening
is what is fully responsible for these shifts". However, it is now
known that there are probable additional photometric effects, caused
by ``multiple stellar populations" in the cluster CMDs, which are
normally more pronounced in both the $U$ passband and $U$-based
colors due to CNO molecular bands located in the UV part of the
spectrum. Also, interstellar extinction is stronger at shorter
wavelengths. Therefore, a variable reddening would results in a
larger scatter in the $U$ magnitudes of stars than in $B,V,$ and
especially in $I$ ones. Thus, the effects caused by the two factors
are entangled and the $U$ magnitude seems to be the most affected.
Interesting enough, it is in the $U$ magnitude that the LFs of the
bump region show very small or even no radial variation.

To quantify and check the noted apparent radial
behavior of the bump, we calculated and used the mean magnitudes
($\overline{U}$, $\overline{U_c}$, $\overline{V}$, $\overline{V_c}$,
and $\overline{I}$) of the isolated bump region in respective
cluster regions and passbands. Strictly speaking, such an estimate
is not fully equivalent to the direct estimate of the bump position,
but the former seems to be more conservative than the latter. The
mean magnitudes and respective r.m.s. values ($\sigma$) were
calculated using relevant command of MIDAS system. We also
calculated the uncertainties in the means ($\sigma_m$), taking into account
the number of stars falling in the defined magnitude range (n and
n$_c$ for the magnitudes uncorrected and corrected for differential
reddening, respectively) of the bump region in each of the three
cluster regions considered. The obtained data are listed in
Table~\ref{table}. They support the apparent behavior of the RGB
bump LFs at different radial distance from the cluster center and
show that: the difference between the mean magnitudes of the bump
region in the outer and inner cluster fields increases with
increasing wavelength. For the uncorrected mean magnitudes, in
particular, it changes from being obviously smaller than the error
of the difference between the means in the $U$ passband and up to
$\Delta\overline{I} =$ 0.13 mag, a factor of 1.5 as large as the
respective error in the $I$ passband. We carried out the same study
and made the same estimates for the GC NGC 6752 using multi-color
cluster photometry by \citet{kravtsovetal14}. The photometric data
for both GCs were obtained using observations gathered with the same
facility. The radial variation of the mean magnitude of bump region
(and bump itself) in NGC 6752 depends
on wavelength in the opposite sense as compared to NGC 3201: it is
insignificant or none in $I$ and the largest in $U$.
\citet{kravtsovetal14} pointed to and showed a notable radial
variation of the RGB bump level in the $U$ magnitude in the sense
that it is brighter in the outer part of NGC 6752. Therefore, the
dependence of the bump level on radial distance in these GCs is in
opposite sense, as well.

Note that photometric effects in the $U$ passband may turn out to be
dissimilar in different data sets in the same GC depending on the
exact response curve of the $U$ passband.
Specifically, the response curve of the standard $U$ passband of the
$UBV$ photometric system should, by definition, encompass the Balmer
jump. For RGs observed from different facilities, it may be difficult
to directly compare $U$-measurements because the strong CNO features
may cause non-negligible photometric effects. This is caused by the
passband modified so that the maxima of their response curves are
approximately 150 - 200{\AA} or even more blue-shifted to exclude
the Balmer jump, as described in 
\citet[and references therein]{kravtsovetal07}. The response curve of 
the $U$ passband used for photometry of NGC 3201 by 
\citet{kravtsovetal09} is close to the standard $U$ passband and it 
is shown in \citet{kravtsovetal14}.

The tendency ({\it we stress that one can only refer to a tendency})
in both the apparent behavior of the RGB bump LFs in different
passbands and the estimates quantifying this behavior disagrees with
the scenario which suggests overall metallicity as a
unique radially varying parameter increasing towards the cluster
center. We stress again the point that the variation in the overall
metallicity at a constant [Fe/H] ratio was achieved by varying the
[O/Fe] ratio. It is unclear how the bump level will depend on overall
metallicity in case of variations in nitrogen or carbon abundances only
or of any combination of the CNO components.
This tendency would agree somewhat better (in $V$ and $I$, but not in $U$)
with decreasing metallicity towards the cluster center. However,
the magnitude of the decreasing [Fe/H] ratio towards smaller $R$
(within PRAD $R \approx$ 9\arcmin) is insufficient to cause a
detectable variation of the bump level, since this variation would be
within the extinction uncertainty. For this reason, we examined
another important contributor to the overall metallicity that could be
responsible for the radial variation of the bump level. 

\begin{figure}
\centerline{
\includegraphics[clip=,angle=-90,width=12.0 cm,clip=]{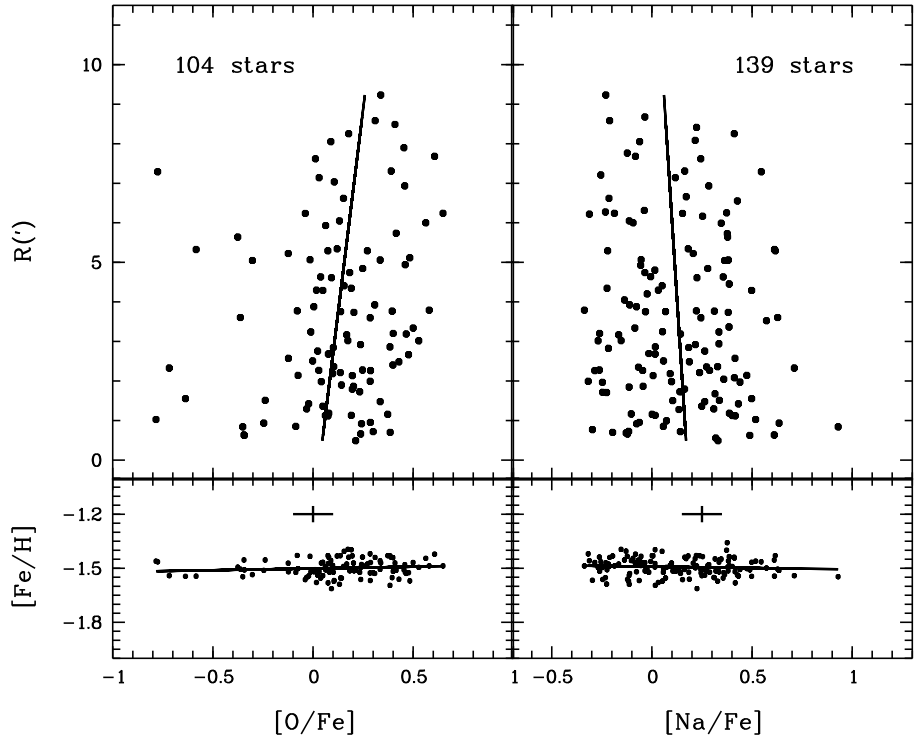}}
\caption{The upper left and right panels show the dependence of
oxygen and sodium abundances in 104 and 139 RGs, respectively, on
the stars' PRAD from the center of NGC 3201, as inferred from the
data of \citet{carrettal09}. The lower panels show relationship
between these abundances and that of iron in the same stars. The
continuous lines are linear fits to the data.} \label{sodoxradep}
\end{figure}

Having analyzed
the data of \citet{carrettal09} on oxygen and sodium abundances in NGC
3201, we found a radial trend of the [O/Fe] ratio in the same sense as
the radial trend of iron abundance, i.e. decreasing oxygen abundance
towards the cluster center. The upper left panel of Fig.~\ref{sodoxradep}
shows the trend in the $R-$[O/Fe] diagram. In turn, the lower left panel
demonstrates an obvious correlation between oxygen and iron abundances
in RGs of NGC 3201. As seen in Fig.~\ref{sodoxradep}, and estimated from
the data, the oxygen abundance variation can be as large as
$\Delta$[O/Fe]$\sim$0.3 dex within PRAD $R \approx$ 9\arcmin.
We computed and analyzed stellar models that mimic the same change in
the [O/Fe] ratio we detected in NGC 3201; we found that the models
predicted a comparable bump-level variation in the $V-$ and $I-$band
simulated data (Fig.~\ref{bumpintracks}).
Also, both the predicted and suspected radial dependencies of the bump
level are of the same sense: the bump magnitude is fainter at higher
[O/Fe] ratio, i.e. at larger PRAD from the cluster center. According to
the same models, the variation of the bump level would be larger in the
$U$ passband and smaller in the $I$ one, contrary to the observed
dependence in NGC 3201. It should be commented in this context that the
two models we rely on are different by only oxygen abundance and the
abundances of other key elements, in particular of carbon and nitrogen,
remain (nearly) unchangeable. As for the real situation in NGC 3201,
there is no guarantee that the radial trend of oxygen (and iron) abundance
is not accompanied by a trend of C or/and N abundances. Indeed,
Fig.~\ref{sodoxradep} shows that there is a radial variation of the
[Na/Fe] ratio of a smaller amplitude and in the sense opposite to
that of both the [O/Fe] and [Fe/H] ratios. Since sodium and nitrogen
abundances are known to correlate with each other in RGs, the radial
trend of the former implies a same-sign trend of the latter (i.e., of
nitrogen abundance) in the cluster. An interesting relationship can
be noted here: the data of \citet{carrettal09} do not reveal any radial
trend of both the [O/Fe] and [Na/Fe] ratios in NGC 6752, where radial
variation of the RGB bump level, as was noted above, is dissimilar
also to that in NGC 3201.

A problematic point should be mentioned. There a lack of agreement
between the RGB bump level variation at different PRAD from the cluster
center, on the one hand, and the appearance of some features of the
CMDs (with different colors and magnitudes) in the same parts of the
cluster, on the other. For example, a fainter bump level (in $V$
and $I$) in the outer than in the inner part of the cluster is in
agreement with higher oxygen and iron abundances (but presumably
a lower nitrogen and carbon abundances, though) in the former than in
the latter, but the shape of the turn-off region and the slope
of the sub-giant branch in the CMDs of the outer cluster part are
apparently correspond more to a more metal-poor population than in
the central part. We do not know anything about the radial behavior
of carbon abundance and of the overall metallicity in NGC 3201. Hence,
a somewhat higher overall metallicity in the central part of NGC 3201
or at least no radial variation of the overall metallicity, cannot be
ruled out.

The real situation within NGC 3201 may be more complicated than the
presence of radially varying abundances of the key elements.
A promising parameter to play a role is radially increasing rotational
velocity of stars in the GC, which ``allows" the stars to be somewhat
more massive at the same age (and even more at younger age) as compared
to the more slowly rotating (and older) counterparts. Comparison of models
with and without rotation \citep{georgetal13} shows that the upper limit
of mass difference between the present-day RGB stars in a GC with overall
metallicity Z$=0.002$ could be as large as $\Delta M \sim 0.05 M_{\sun}$
under an assumption that rotating models (initial equatorial velocity on the
zero age main sequence is $\sim10$ km s$^{-1}$ for the initial mass around
$M_{in} = 0.80 M_{\sun}$), younger age, within 1 Gyr, and
the same helium abundance. Increasing helium abundance in rotating RGB
stars should decrease $\Delta M$. The effect of slightly increased
$\Delta$Y (by $\sim 0.01 - 0.02$) on $\Delta M$ can be compensated by a
somewhat higher ($\sim0.15$ dex) overall metallicity. We thus conclude
that the largest value of $\Delta M$, even somewhat larger than
$0.05 M_{\sun}$, can {\it formally} be achieved in a GC if rotating RGB
stars are more metal rich and younger but with the same helium abundance
as compared to their non-rotating counterparts. Assuming the existence 
of two distinct sub-populations in NGC 3201, \citet{wagkaisetal16} 
estimated (upper limit of) helium difference among them around 
$\Delta$Y$= 0.06$. If real, such fairly large difference in helium 
abundance would rule out the possibility of increased mass of the 
centrally located sub-population of stars in the cluster.
At this point, two comments should be made. First, the lowest metallicity
(Z$=0.002$) of the available rotating models is higher than the metallicity
of NGC 3201 and, strictly speaking, $\Delta M$ may be smaller at lower metallicity.
On the other hand, the real difference (if any) between typical initial
rotational velocities of stars (of a given initial mass) of the two assumed
sub-populations is unknown quantity and it could be higher than the unique option
available for the rotating models. Moreover, we do
not have reliable information about the real (radial) variation of the overall
metallicity in NGC 3201. Second, the rotating models do not unfortunately fully
trace the RGB but only its lower part and do not even achieve the level
of the RGB bump. These limitations impose serious constraints on or
exclude a more complete analysis.

A higher metallicity RG can have a higher mass than a comparable star of lower
metallicity. In turn, both a faster rotation and a higher mass make the RGB and
its bump brighter (and hotter) partially compensating for the
opposite effect of increasing metallicity (if any) on the temperature
and luminosity towards the cluster center or making the bump brighter at
a constant metallicity.
On the other hand, the same-sign radial trends of the mass and rotational
velocity will result in opposite-sign (compensating each other) effects
on the rate of stellar evolution, allowing thereby a RGB star to be more
massive than its more metal-poor and slowly rotating counterparts at the
same point in their evolution. Unfortunately, the available models with
rotation do not allow a study of this subject in more detail, to estimate
the variation of both the luminosity and temperature of the RGB bump as a
function of metallicity and rotational velocity. However, we refer to
\citet{brownsalar07} who estimated quantitatively (for a given set of age,
metallicity, etc.) the impact of rotation on increasing brightness of the
RGB bump. It should be stressed that the combined effect of the parameters
under consideration on the RGB stars and on the bump level must be in
agreement with the respective effect of the same parameters on the
horizontal branch stars. It is pertinent to note in the context of our
present study that \citet{tailoetal15} arrive at conclusion of rapidly
rotating second generation stars in the GC $\omega$ Cen.

Considering the variation of the He abundance, at the same age and the
same range of the rotational velocity of GC stars, $\Delta M$ should
be lower or even zero, depending on $\Delta$Y. The possibility of a
detectable mass difference among RGB stars in GCs with a (large)
variation of the N
abundance is implied by very recent result on the field RGB stars:
\citet{martigetal16} found a tight correlation between the mass of RGB
stars and the ratio of their N and C abundances at a given [Fe/H],
$T_\mathrm{eff}$, and Log $g$. In the context of GC multiple
populations, is there any variation in the the N/C ratio at a given
$T_\mathrm{eff}$, and Log $g$ among RGB stars within separate monometallic
GCs? There is need of further study of possible mass, rotational velocity, and N/C
ratio variations among RGB stars in GCs \citep[see][]{grattonetal12}.

Note that increasing mass and/or increasing
rotational velocity, and/or increasing helium abundance of stars
toward the GC centers are in agreement with colors getting bluer
toward the same direction.

To complete the discussion, we summarize the results derived from 
the older data sets: (1) H$\alpha$ emission is present in the
spectra of four sample stars of \citet{gonzawaller98}; (2) the mean 
[Fe/H] value of these RGs is somewhat lower than for the rest of the 
stars; (3) the most metal-poor sample star, C\^{o}t\'{e} 246, shows 
the strongest H$\alpha$ emission.

\section{Conclusions}
\label{conclus}

We studied the unusual radial trend of the iron abundance in the GC
NGC 3201. It was originally found, at statistically significant
level in some GCs including NGC 3201, as a slightly decreasing
[Fe/H] ratio in red giants towards the GC centers, by relying on the
data of \citet{carrettal09} alone.

We discovered if the trend is reproduced by distinct data on iron
abundance independently derived for the same GC. We compared four
sets of data obtained by different authors using observations
gathered with different facility and/or reduced independently, but
all the data are dominated by those of \citet{carrettal09} until
more data can be gathered for distant RGB stars.

The data of \citet{simmereretal13}, based on FLAMES-UVES and MIKE
spectra, also reveal the trend of the same sense (more centrally
concentrated LIA RGs) at marginally statistically significant level.
\citet{mucciaretal15} re-analyzed the same FLAMES-UVES spectra and
reported on negligible spread of iron abundance derived from FeII
spectral lines. We find that these data support the trend at a high
confidence level of 92\%. The trend is virtually indistinguishable,
within the error, form that originally found using much larger
sample size of \citet{carrettal09}'s data. Additionally, we notice
that both data sets are indistinguishable from each other with
respect to their standard deviations giving evidence of the same
scatter in the data. Revising the present paper we additionally
considered the data of \citet{gonzawaller98} on iron abundance
derived for eighteen RGs. Although this sample size is smaller than
those of other considered data sets, five sample stars were observed
and studied at unprecedentedly long radial distance from the cluster
center ever reached in other studies in NGC 3201. Interesting
enough, these RGs are obviously more iron-rich than the rest of the
stars, thereby supporting (at least formally) the presence of the
trend under consideration. According to \citet{gonzawaller98}, an
effect of a systematic error cannot be excluded, though. Apart from 
the apparent trend in [Fe/H], we found evidence for a more pronounced 
trend in oxygen abundance towards the cluster center. The variation of the
[O/Fe] ratio in RGs as a function of radial distance from the
cluster center can be as large as $\Delta$[O/Fe]$\sim$0.3 dex within
PRAD $R \approx$ 9\arcmin.

We considered and discussed a number of causes  and effects that might
mimic the radial variation of iron abundance. The
same effects could, however, hardly mimic the notable trend of
oxygen abundance as well, taking into account the presence of a
similar radial trend of the [Fe/H] ratio in the GC NGC 6752 but no
apparent oxygen abundance trend. These effects might be related to the
physical characteristics of the studied objects themselves, in
particular, to a faster rotation and a higher mass of centrally
located RGs. Both these parameters should make RGs to be somewhat
hotter and brighter than their less massive and more slowly rotating
counterparts (at least at the same overall metallicity and helium
abundance), but simultaneously they have opposite, mutually
compensating effects on the time scale of stellar evolution. A younger age
of rotating stars strengthens these effects. The RGB bump
LFs show a tendency for the bump to became brighter towards the
cluster center in the $V$ and $I$ (but presumably not in $U$) passbands.
However, the radial trend of the rotational velocity and mass of
RGs in NGC 3201 could hardly be responsible for the estimated magnitude
of the brightening of the bump level towards the cluster center. It
could not be the main contributor to the brightening. Several radially
dependent photometric effects can
alter the appearance of CMDs with different magnitudes and colors that
contradict each other and need to be reconciled. Partially, this can
be caused by the combined effects of ``multiple populations" and
irregularly varying reddening. An increasing nitrogen abundance toward
the cluster center is possible. Also, radially increasing carbon
abundance in the same sense cannot be ruled out, since the abundances
of these two elements are known to correlate with each other in GCs.
Helium abundance slightly increasing (by $\Delta$Y $\sim 0.01 - 0.02$)
towards the cluster center cannot be excluded, as well.

Stellar rotation is an important parameter poorly considered in present-day 
studies of GCs. An increased rotation of more centrally located RGs in GCs, 
especially in
combination with their higher metallicity and younger age, could potentially
lead to a non-negligible radial variation of mass, increasing toward
the cluster center, among RGs of the same evolutionary stage. Another
alternative may be that NGC 3201 might be of notably higher mass in
the past and the primordial radial effects in the initial GC might
partially be conserved. Also, tidal stripping removes lower-mass stars
first and concentrates higher mass stars and binaries in the core.
This GC is well-known by its unusual kinematics in the Galaxy
\citep{harris96} and the origin of NGC 3201 is not clear.

Finally, the analysis presented in our study shows that the bulk of 
the data on iron abundance, in particular, and on elemental abundances, 
in general, in NGC 3201 were obtained for stars located within radial
distance $R \approx$ 9\arcmin.\  There is a lack of reliable
information and therefore intriguing uncertainty about radial
dependence of iron (elemental) abundance(s) at (much) larger radial
distance from the cluster center, since the measurements made by
\citet{gonzawaller98} for five stars in the range 13\arcmin $< R <$
26\arcmin\ from the cluster center are the unique data obtained in
the outer(most) part of NGC 3201.

\acknowledgements The author thanks the anonymous referee for useful
comments and suggestions that improved the manuscript. This research
has made use of the VizieR catalogue access tool, CDS, Strasbourg, France.

\clearpage

\clearpage

\clearpage

\clearpage

\clearpage

\clearpage

\clearpage

\clearpage

\clearpage

\clearpage

\begin{deluxetable}{cllllllllllllllll}
\tabletypesize{\scriptsize} \tablecaption{Mean magnitudes of the RGB
bump region at different radial distance. \label{table}}
\tablewidth{0pt} \tablehead{ \colhead{Area\tablenotemark{a}}  &
\colhead{n; n$_c$} & \colhead{$\overline{U}$} & \colhead{$\sigma_m$}
& \colhead{$\sigma$} & \colhead{$\overline{U_c}$} &
\colhead{$\sigma_m$} & \colhead{$\sigma$} & \colhead{$\overline{V}$}
& \colhead{$\sigma_m$} & \colhead{$\sigma$} &
\colhead{$\overline{V_c}$} &  \colhead{$\sigma_m$}&
\colhead{$\sigma$} & \colhead{$\overline{I}$} & \colhead{$\sigma_m$}
& \colhead{$\sigma$} } \startdata
inn & 42; 40& 16.29 & 0.04 & 0.28 & 16.34 & 0.04 & 0.25 & 14.77 & 0.06 & 0.36 & 14.79 & 0.05 & 0.34 & 13.50 & 0.06 & 0.39\\
int & 41; 40& 16.34 & 0.05 & 0.29 & 16.35 & 0.05 & 0.28 & 14.82 & 0.06 & 0.38 & 14.81 & 0.06 & 0.38 & 13.55 & 0.06 & 0.40\\
out & 45; 43& 16.27 & 0.04 & 0.26 & 16.37 & 0.04 & 0.24 & 14.87 & 0.05 & 0.34 & 14.94 & 0.05 & 0.33 & 13.63 & 0.06 & 0.37\\
\enddata
\tablenotetext{a}{The "inn"-, "int"-, and "out"-areas denote those cluster regions, radially getting away from the cluster center, the RGB bump LFs of which are shown in the upper, intermediate, and lower panels of  Fig.~\ref{bumpraddep}, respectively.}

\end{deluxetable}

\end{document}